\pdfoutput=1
\documentclass[pdflatex,sn-mathphys-ay]{sn-jnl}
\usepackage{amsmath,amssymb}
\usepackage[T1]{fontenc}
\usepackage[utf8]{inputenc}
\usepackage{microtype}
\usepackage{booktabs}
\usepackage{graphicx}
\usepackage{tikz}
\hypersetup{hidelinks}

\numberwithin{equation}{section}

\begin{document}

\title[The two-component picture of soft hadron production]{The two-component picture of soft hadron production: Regge exchange, duality, and resonance cascades}

\author*[1,2]{\fnm{Ondrej} \sur{Chvala}}\email{ochvala@utexas.edu}

\affil*[1]{\orgdiv{Walker Department of Mechanical Engineering}, \orgname{The University of Texas at Austin}, \orgaddress{\city{Austin}, \state{TX}, \country{USA}}}

\affil[2]{\orgdiv{Department of Nuclear Engineering}, \orgname{The University of Tennessee}, \orgaddress{\city{Knoxville}, \state{TN}, \country{USA}}}

\abstract{The fragmentation regions of soft hadronic collisions approach
energy-independent limits. Benecke, Chou, Yang, and Yen stated this
hypothesis of limiting fragmentation in 1969, and Feynman gave it a
parton-model formulation the same year. Measurements from the CERN
Intersecting Storage Rings (ISR) to the Large Hadron Collider (LHC)
show this approach. Fischer and co-workers interpret the NA49 spectra from the CERN Super Proton Synchrotron (SPS) as the sum of a target and
a projectile component, each the decay of an excited hadronic system.
This review traces the theory behind that two-component picture. It
examines the hypothesis that the hadronic tools that predate quantum
chromodynamics (QCD) suffice for the soft, non-perturbative domain of
QCD. Regge exchange organizes the fragmentation region in triple-Regge
form and supplies the diffractive mass law of the excited systems.
Finite-energy sum rules and the two-component duality of Harari and
Freund equate averaged towers of resonances with that exchange. Duality
therefore motivates describing the excited systems in hadronic degrees
of freedom. The cascade description has five
predecessors, namely the fireballs of Niu, the novas of Jacob and
Slansky, the clusters of Hwa, Good--Walker diffraction, and the
two-stage dissociation of Hama. A companion paper (arXiv:2611.11570)
assembles these pieces into HEART (Hadronic Excitations And Resonance
Transitions), an event-level cascade of two-body hadronic transitions,
and fits it to the NA49 data. The review closes with the
fragmentation-region measurements that would separate a resonance
cascade from a string, namely the isospin of forward baryons, local strangeness
balance, and forward resonance signals.}


\maketitle

\section{Introduction}\label{sec:intro}

Most collisions between two protons at high energy are soft. The
protons exchange little transverse momentum, and the hadrons they
produce carry transverse momenta of a few hundred MeV/$c$. These soft
hadrons make up the bulk of the particles in every such collision. The
theory of the strong interaction, quantum chromodynamics (QCD),
describes quarks and gluons whose coupling weakens at short distances.
Perturbative calculations in that coupling therefore apply to hard
collisions, in which a large momentum transfer resolves short
distances. Soft collisions probe distances of the size of the proton
itself, where the coupling is strong and no perturbative expansion
holds. The production of the soft bulk of hadrons has therefore not yet
been calculated from first principles.

In the 1950s and 1960s, before QCD, physicists described soft
production with a toolkit built on general properties of the
scattering amplitude. Unitarity expresses the conservation of
probability. Analyticity makes the amplitude a smooth function of
energy and angle. Isospin symmetry treats the proton and the neutron
as two states of one particle. Regge theory related the energy
dependence of a reaction to the families of hadrons that can be
exchanged in it~\citep{regge1959complex}. Duality equated an average
over the resonances formed in a reaction with that
exchange~\citep{dolenhornschmid1968,harari1968pomeranchuk}. Fireball
and statistical models described the produced hadrons as the decay
products of excited, short-lived
objects~\citep{niu1958fireball,hagedorn1965thermo}. After the
discovery of asymptotic freedom in
1973~\citep{grosswilczek1973,politzer1973}, the attention of the
field moved to hard processes, where QCD makes precise predictions.

This review follows one line of these ideas from their origin to the
present. It starts from a regularity of the measured spectra in the
fragmentation regions, the parts of phase space that hold the hadrons
moving close to either beam. The review then follows the ideas that
explain this regularity, in the order in which they build on each
other, up to an event-level model of today.
Figure~\ref{fig:timeline} places the steps in time, from 1958 to 2026.
Every term is defined at first use, and
Appendix~\ref{sec:nomenclature} collects the terms.

\begin{figure*}[htbp]
  \centering
  \begin{tikzpicture}[>=latex]
    \draw[thick] (0,0) -- (0,-14.4);
    \tikzset{yr/.style={anchor=east, font=\small},
             ev/.style={anchor=west, text width=10.5cm, font=\small}}
    \foreach \y/\t in {
      0/1958--59, 0.9/1960, 1.8/1965, 2.7/1968, 3.6/1969, 4.5/1970,
      5.4/1971, 6.3/1972, 7.2/1975, 8.1/1982--83, 9.0/1992, 9.9/1998,
      10.8/2003--2010, 11.7/2009--2013, 12.6/2019, 13.5/2022, 14.4/2026} {
      \draw (0,-\y) -- (-0.15,-\y);
      \node[yr] at (-0.25,-\y) {\t};
    }
    \node[ev] at (0.3,0) {Niu two-center model; Regge complex angular momenta};
    \node[ev] at (0.3,-0.9) {Good--Walker diffractive eigenstates};
    \node[ev] at (0.3,-1.8) {Hagedorn statistical thermodynamics of strong interactions};
    \node[ev] at (0.3,-2.7) {Finite-energy sum rules; two-component duality; Veneziano amplitude};
    \node[ev] at (0.3,-3.6) {Benecke et al.\ limiting fragmentation; Feynman parton frame};
    \node[ev] at (0.3,-4.5) {Mueller generalized optical theorem; DeTar--Snider multiperipheral account};
    \node[ev] at (0.3,-5.4) {Jacob--Slansky pion distributions; Hwa clusters; triple-Regge spectra};
    \node[ev] at (0.3,-6.3) {Jacob--Slansky nova model (full paper); Hama two-stage dissociation; inclusive duality; two-pion confirmation};
    \node[ev] at (0.3,-7.2) {Fuchi--Niu two-fireball model at accelerator energies};
    \node[ev] at (0.3,-8.1) {ISR two-particle correlations; Goulianos diffractive review};
    \node[ev] at (0.3,-9.0) {Donnachie--Landshoff soft Pomeron};
    \node[ev] at (0.3,-9.9) {Collins: factorization fails in diffractive hadron-hadron collisions};
    \node[ev] at (0.3,-10.8) {Two-component picture of Fischer et al.; NA49 p+p spectra};
    \node[ev] at (0.3,-11.7) {Gustafson, Good--Walker, and triple-Regge under Koba--Nielsen--Olesen scaling};
    \node[ev] at (0.3,-12.6) {Limiting fragmentation tested at the LHC};
    \node[ev] at (0.3,-13.5) {Fischer global interpolation of $\pi^-$ from 3~GeV to 13~TeV};
    \node[ev] at (0.3,-14.4) {HEART model as event-level realization};
  \end{tikzpicture}
  \caption{Timeline of the ideas this review traces, from the two-center
  model through the NA49 spectra and the 2022 interpolation. Dates are  publication years.  Full citations are in the References.}
  \label{fig:timeline}
\end{figure*}

In 1969 Benecke, Chou, Yang, and Yen stated that the fragmentation
regions of soft hadronic collisions approach energy-independent
limits~\citep{benecke1969limiting}. Feynman framed the same limits in
parton language that year~\citep{feynman1969veryhigh}. Fischer and
co-workers interpreted the NA49 spectra in p+p collisions at 158~GeV/$c$ as
the sum of a target component and a projectile component, each the
decay of an excited hadronic system
\citep{varga2003baryontransfer,chvala2004isospin,chvala2013backward}.
Fischer, Makariev, Varga, and Wenig built a global interpolation of the
$\pi^-$ cross sections on that picture, from $\sqrt{s} = 3$~GeV to
13~TeV \citep{fischer2022piminus}.

Below $p_T \approx 1.3$~GeV/$c$ hadron production lies in the soft,
non-perturbative domain of QCD.
Analyticity, unitarity, and isospin hold there as properties of the
$S$-matrix. Regge exchange and resonance duality remain applicable
there as phenomenological frameworks. This review examines the
hypothesis that these tools suffice for that domain. In that
hypothesis each experimental component is the decay of one excited
hadronic system, and the resonances of that system build the Regge
exchange that produced it. A tower of resonances and a string are then two descriptions of the same excited object.
Section~\ref{sec:measure} names the fragmentation-region observables
on which these two descriptions separate.

This review is written for the physicist who wants the soft bulk of the strong interaction described with completeness comparable to that of the hard regime. The field's effort has moved to hard scattering and perturbative QCD, and the hadronic toolkit of the soft heart has fallen out of everyday use. This review collects that toolkit as background, from the energy-independent fragmentation limits through Regge exchange and the duality bridge to the cascade realizations. The verdict between realizations belongs to measurements.

The companion paper defines HEART and fits it to the published NA49
cross-section data~\citep{heartpaper2026}. This review carries the
theory that the model uses, one layer at a time. The NA49 spectra are
taken as published. 

The paper is arranged as follows.
Section~\ref{sec:lf} states limiting fragmentation. 
Section~\ref{sec:domain} describes the domain in which these tools apply. Section~\ref{sec:regge} organizes the fragmentation region in triple-Regge form. It derives the triple-Pomeron law with the transfer dependence that the companion paper's model takes. 
Section~\ref{sec:duality} states duality as the bridge between resonance towers and Regge exchange.
Section~\ref{sec:precursors} traces the cascade through its predecessors and introduces HEART's continuum density and decay weight beside them. Section~\ref{sec:heart} assembles the four layers of HEART from those pieces. It also gives one schematic event and a dictionary of Regge-era terms in event-generator language. 
Section~\ref{sec:measure} names the measurements that separate the cascade from a string.
Section~\ref{sec:conclusions} concludes, with the open questions in Section~\ref{sec:outlook}.
Appendix~\ref{sec:nomenclature} collects the terms used.

\section{Limiting fragmentation}\label{sec:lf}

The limiting-fragmentation hypothesis states that the fragments of a
beam particle, viewed in that particle's rest frame, become independent
of the beam energy as the energy tends to infinity. The 1969 formulation of
Benecke, Chou, Yang, and Yen is the sharpest:
\begin{quote}
  \emph{In a frame where one of the beam particles is at rest, the
  distribution of fragments of that particle becomes independent of the
  energy of the incident beam as the energy tends to infinity}
 ~\citep{benecke1969limiting}.
\end{quote}
In invariant form, the invariant cross section in the fragmentation
region approaches a scaling function of $x_F$, $p_T$, and the produced
hadron species $h$,
\begin{equation}\label{eq:lf_limit}
  E\frac{d^3\sigma}{dp^3}
  \xrightarrow[\sqrt{s}\to\infty]{} F_h(x_F, p_T)\,,
\end{equation}
at fixed $x_F$ and $p_T$. The left side is the invariant cross section $f \equiv E\,d^3\sigma/dp^3$, the quantity tabulated in the published NA49 cross-section data. The 1969 formulations took the interaction cross sections as constant, so they did not separate the invariant cross section from the invariant density per inelastic event, $(1/\sigma_{\mathrm{inel}})\,E\,d^3\sigma/dp^3$. At finite energy the two differ, since $\sigma_{\mathrm{inel}}$ rises by about 20\% between $\sqrt{s} = 3$ and 63~GeV~\citep{fischer2022piminus}. 
Feynman stated the same limit in terms of rapidity. As the center-of-mass energy increases, the rapidity distribution extends in both directions, while the distribution in the rest frame of either beam tends to a limiting shape~\citep{feynman1969veryhigh}.

\begin{figure*}[htbp]
  \centering
  \includegraphics[width=0.85\textwidth]{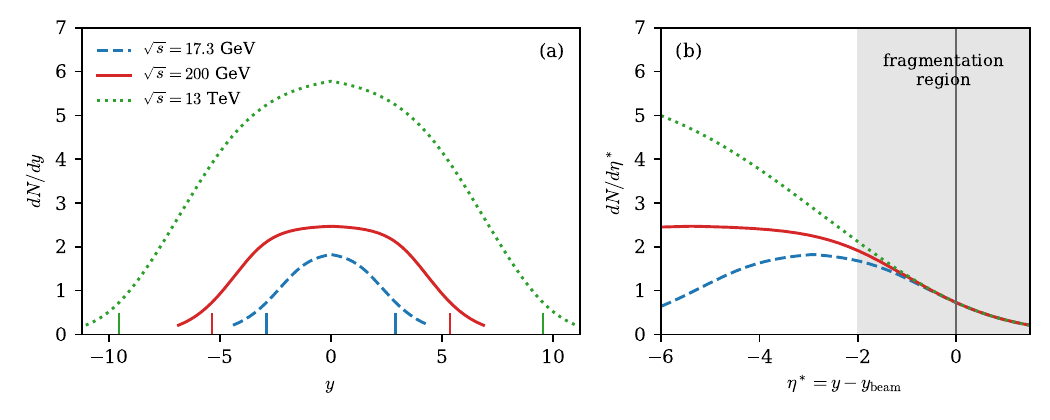}
  \caption{Schematic of limiting fragmentation. The curves are
  illustrative and show no data. Their central heights are near the
  p+p values.
  (a)~Rapidity densities $dN/dy$ at $\sqrt{s} = 17.3$~GeV (broken
  line), 200~GeV (full line), and 13~TeV (dotted line). The short
  ticks mark $\pm y_{\mathrm{beam}}$ of each energy as reference
  rapidities. The central region rises and widens with energy.
  (b)~The projectile half of the same curves against the beam-frame
  rapidity $\eta^* = y - y_{\mathrm{beam}}$. In the fragmentation
  region, shaded, the three curves coincide. Toward more negative
  $\eta^*$ each curve passes into its own central region, which
  begins farther from the beam at higher energy.}
  \label{fig:lim_frag_schematic}
\end{figure*}

For $A + B \to h + X$ the kinematics are
\begin{align}
  s    & = (p_A + p_B)^2 \,, \\
  x_F  & \equiv \frac{2p_L^*}{\sqrt{s}} \,, \\
  y_h  & = \frac{1}{2}\ln\frac{E_h + p_L^*}{E_h - p_L^*}
          = \mathrm{arcsinh}\left(\frac{p_L^*}{m_T}\right) \,,
\end{align}
with starred quantities in the center-of-mass frame and $m_T = \sqrt{m_h^2 + p_T^2}$ the transverse mass. The beam rapidities are $y_A^{\mathrm{beam}} \approx +\ln(\sqrt{s}/m_A)$ and $y_B^{\mathrm{beam}} \approx -\ln(\sqrt{s}/m_B)$, up to mass corrections. For the projectile $A$ we write $y_{\mathrm{beam}} \equiv y_A^{\mathrm{beam}}$. The fragmentation region of $A$ is a finite interval of rapidity adjacent to $y_{\mathrm{beam}}$. The convenient variable is the beam-frame rapidity $\eta^* \equiv y - y_{\mathrm{beam}}$. Figure~\ref{fig:lim_frag_schematic} sketches the limit in this variable. In this variable the hypothesis states that $dN/d\eta^*$ at fixed $\eta^*$ becomes independent of $\sqrt{s}$ near $\eta^* = 0$, over an interval about two units wide. The statement concerns the density per inelastic event, under the constant interaction cross sections of the 1969 formulations. Light secondaries also reach small positive $\eta^*$, just beyond the beam rapidity, because their mass is small against the beam nucleon's. Fischer and collaborators use the reduced rapidity $y_{\mathrm{lab}} = y_{\mathrm{beam}} - y$, so that $\eta^* = -y_{\mathrm{lab}}$. The projectile fragmentation region is then the region of small $y_{\mathrm{lab}}$. At $\sqrt{s} = 17.3$~GeV the beam rapidity is $y_{\mathrm{beam}} = 2.91$. The interval $-2 < \eta^* \le 0$ then corresponds to $0.91 < y \le 2.91$ in the center-of-mass frame. Toward more negative $\eta^*$ the hadron moves away from the projectile, through mid-rapidity at $\eta^* = -2.91$ into the target hemisphere.

This review states hypotheses in rapidity and quotes numbers in $x_F$. The scaling statement of Equation~\eqref{eq:lf_limit} uses $x_F$, and experiments tabulate in it. Positions, shapes, and widths are boost-invariant only in rapidity. At fixed transverse mass the two variables carry the same information. For a forward-moving hadron with $\sqrt{s} \gg m_N$, where $m_N$ is the beam nucleon mass, the definitions above give
\begin{equation}\label{eq:xf_vs_etastar}
  x_F \approx \left(m_T/m_N\right) e^{\eta^*} \,.
\end{equation}
The position in $\eta^*$ depends on $x_F$ and $m_T$ together. At fixed $x_F$ a lighter hadron lies closer to the beam rapidity. A pion at $x_F = 0.2$ with $m_T/m_N = 0.3$ lies at $\eta^* \approx -0.4$, a proton at the same $x_F$ and $p_T = 0.3$~GeV/$c$ at $\eta^* \approx -1.6$, and the leading proton at $x_F$ near one at $\eta^* \approx 0$. Near the kinematic limit at the energy of the CERN Super Proton Synchrotron (SPS), Fischer and collaborators use the variable $x_F' = p_L/p_L^{\max}$ in place of $x_F$~\citep{fischer2022piminus}.

\begin{figure*}[htbp]
  \centering
  \begin{tikzpicture}[scale=1.0,>=latex]
    \fill[gray!20] (-0.8,0) rectangle (0.8,2.9);
    \node at (0,2.65) {\small overlap};
    \draw[->] (-3.6,0) -- (3.6,0) node[below] {$y$};
    \draw[->] (-3.6,0) -- (-3.6,3.2) node[left] {$E\,d^3\sigma/dp^3$};
    \draw[thick,dashed,blue] plot[smooth] coordinates
      {(-3.5,0.12) (-2.5,1.2) (-1.5,2.2) (-0.5,1.45) (0.3,0.55) (1.0,0.12)};
    \draw[thick,dotted,red] plot[smooth] coordinates
      {(3.5,0.12) (2.5,1.2) (1.5,2.2) (0.5,1.45) (-0.3,0.55) (-1.0,0.12)};
    \draw[thick,black] plot[smooth] coordinates
      {(-3.5,0.15) (-2.5,1.35) (-1.5,2.5) (-0.5,2.25) (0,2.2)
       (0.5,2.25) (1.5,2.5) (2.5,1.35) (3.5,0.15)};
    \node[blue] at (-2.75,2.6) {\small target};
    \node[red] at (2.75,2.6) {\small projectile};
    \node at (0,1.5) {\small sum};
  \end{tikzpicture}
  \caption{Schematic of the two-component picture after Fischer and
  co-workers. The target component (dashed) and the projectile
  component (dotted) each peak in their own hemisphere and extend into
  the other. Where they overlap, the measured yield is their sum
  (solid). The overlap width is $\Delta\eta \sim 1$. The figure shows
  no data, and its shapes and normalization are illustrative.}
  \label{fig:two_component}
\end{figure*}
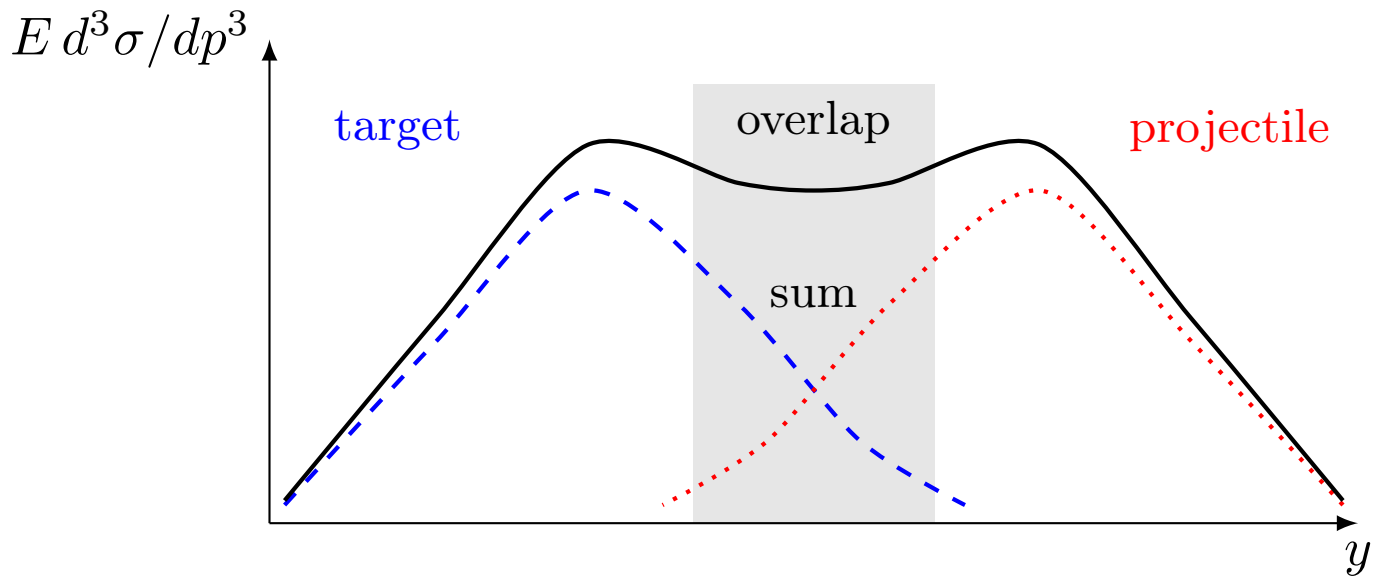

Table~\ref{tab:exp_status} collects three experimental tests.
Equation~\eqref{eq:lf_limit} states the energy independence of the invariant cross section. 
The row of the CERN Intersecting Storage Rings (ISR) tests a different statement. The R-607 spectrometer found no correlation between projectile and target secondaries at $|x_F| > 0.2$ in p+p collisions at $\sqrt{s} = 44.7$ and 62.3~GeV. That absence of correlation is consistent with a weak coupling of the selected fragments of the two hemispheres~\citep{Bobbink1982}. 
The baryon-transfer analyses of Varga, Chvala, and co-workers put the target-projectile overlap (``feed-over'') at $\Delta\eta \sim 1$ in beam-rapidity units~\citep{varga2003baryontransfer,varga2004baryonpair,chvala2004isospin}. Figure~\ref{fig:two_component} sketches the two components and their overlap.
Fischer and collaborators test limiting fragmentation on their 2022 global interpolation of the $\pi^-$ cross sections, which spans $3 < \sqrt{s} < 63$~GeV~\citep{fischer2022piminus}. They work in the target rest frame, at negative longitudinal momentum $p_L^{\mathrm{lab}}$, the region in which the 1969 formulation states the limit for pions. There the double-differential cross section $d^2\sigma/(dp_L^{\mathrm{lab}}\,dp_T)$ agrees closely over the whole energy range. The cross section per inelastic event varies more with energy, since $\sigma_{\mathrm{inel}}$ rises~\citep{fischer2022piminus}.
The same comparison reaches the Large Hadron Collider (LHC) in the very forward region. The LHCf experiment measured the $\pi^0$ cross sections $d^2\sigma/(dp_L^{\mathrm{lab}}\,dp_T)$ at $\sqrt{s} = 2.76$ and 7~TeV. They agree with the prediction of the interpolation from NA49 and ISR energies, for $p_T \gtrsim 0.2$~GeV/$c$ at 7~TeV and at all $p_T$ at 2.76~TeV. The densities per inelastic event, $f/\sigma_{\mathrm{inel}}$, differ between the two energies~\citep{fischer2022piminus}. For charged hadrons in central Au+Au and Pb+Pb collisions, Kellers and Wolschin extrapolate the measured pseudorapidity distributions into the fragmentation region, where no LHC data exist. They find them compatible with limiting fragmentation from 19.6~GeV to 5.02~TeV~\citep{Kellers2019}. That result is a model extrapolation, so Table~\ref{tab:exp_status} does not list it among the tests.

\begin{table*}[htbp]
  \centering
  \caption{Experimental status of limiting fragmentation. The ISR row
  is the absence of correlation between projectile and target
  secondaries at $|x_F| > 0.2$, a test of the coupling of the selected
  fragments~\citep{Bobbink1982}.
  The fixed-target row tests energy independence of the $\pi^-$ cross
  section $d^2\sigma/(dp_L^{\mathrm{lab}}\,dp_T)$ in the target rest
  frame at $p_L^{\mathrm{lab}} < 0$, on the 2022 global interpolation of
  the bubble-chamber, NA49, and ISR data~\citep{fischer2022piminus}. The
  LHC row compares the very forward $\pi^0$ cross sections of LHCf with
  the prediction of that interpolation from the lower
  energies, at $p_T \gtrsim 0.2$~GeV/$c$ for 7~TeV and at all $p_T$ for
  2.76~TeV~\citep{fischer2022piminus}.}
  \label{tab:exp_status}
  {\footnotesize\setlength{\tabcolsep}{4pt}
  \begin{tabular}{@{}lcccc@{}}
    \toprule
    Experiment  & $\sqrt{s}$    & System    & Region        & Result      \\
    \midrule
    ISR         & 44.7--62.3 GeV & p+p       & pairs at $|x_F| > 0.2$ & no $x_F$ correlation \\
    Fixed target to ISR & 3--63 GeV & p+p, $\pi^-$ & $p_L^{\mathrm{lab}} < 0$ & energy-independent $d^2\sigma$ \\
    LHCf        & 2.76--7 TeV   & p+p, $\pi^0$ & very forward & energy-independent $d^2\sigma$ \\
    \bottomrule
  \end{tabular}}
\end{table*}

In the adopted Regge description the deviations from the limit take
known forms. Subleading Reggeon contributions enter as
$F_1(x_F)/s^{\alpha}$ with $\alpha \sim 0.5$. The energy factor
$s^{\epsilon}$ of the triple-Pomeron term (Section~\ref{sec:regge})
lets the invariant cross section rise slowly at fixed $x_F$. The
rising small-$x$ gluon density feeds particle production away from
the fragmentation peak. Its effect on the fragmentation region itself
is beyond the scope of this review. So is the logarithmic Dokshitzer--Gribov--Lipatov--Altarelli--Parisi
(DGLAP) evolution of fragmentation functions. That evolution resums
logarithms of a hard scale far above the hadronic one. Below
$p_T \approx 1.3$~GeV/$c$ every momentum in the process is of hadronic
size, so no such logarithm arises (Section~\ref{sec:domain}).

\subsection{Feed-over and feed-down: overlap and weak-decay conventions}

Two data-treatment words recur in this review because they decide
which tables a cascade may be compared to. \emph{Feed-over} is the
target-projectile overlap. Each component extends past mid-rapidity
into the other's hemisphere, so the measured yield near $x_F = 0$ is
a sum. Varga, Chvala, and co-workers quantified it in the NA49 data
with baryon-free projectiles and tagged forward baryons. The baryon stopping resides in the projectile component, while the target component does not depend on the projectile. The overlap is a strip $\Delta\eta \sim 1$ wide in beam
rapidity
\citep{varga2003baryontransfer,varga2004baryonpair,chvala2004isospin}.
Outside the strip each hemisphere holds one component. Within about
two units of its beam rapidity that component is the fragmentation
region of its own beam particle.

\emph{Feed-down} is weak-decay subtraction. The detectors see stable
hadrons. A share of the pions and protons comes from weak decays, such
as $\pi$ from $K^0_S$ and $p$ from $\Lambda$, with further contributions from $\Sigma$ and $\Xi$ decays. The published NA49 cross-section data
compared in the companion paper are feed-down subtracted. The yields of the $V^0$ decays, the neutral weak decays of $K^0_S$ and $\Lambda$ seen as a V-shaped pair of tracks, were
measured in the same
detector and their decay products removed from the charged-hadron
spectra~\citep{na49pppi2006}. Data without this subtraction are called feed-down included. A model compared with these tables must follow the same
convention. HEART's event record leaves $K^0_S$, $K^0_L$, $\Lambda$, $\Sigma$, and
$\Xi$ stable. Its observables then perform the decays the tables keep,
namely $\Sigma^0 \to \Lambda\gamma$ and the $\Xi$ and $\Omega^-$
decays, with their antiparticles. The decays the tables subtract stay
undone, so the model and the data follow one feed-down convention. A
model that decays all the hyperons must be compared to
feed-down-included tables instead. A comparison across the two conventions carries a normalization error of the size of the feed-down share.

\section{The domain argument}\label{sec:domain}

The published NA49 cross-section data used in this review span $0.05 \le p_T \le 1.5$~GeV/$c$. Fischer and collaborators take $p_T < 1.3$~GeV/$c$ as the non-perturbative sector~\citep{fischer2022piminus}.
At transverse momenta of that order, the strong coupling is near one half, so there is no small parameter to expand in and no large logarithm of scale ratios to resum. The virtualities of the would-be hard subgraph lie in the hundreds of MeV, the range of the hadronic matrix elements. No Wilson coefficient can therefore be computed and split off. Short-distance scattering and long-distance hadronization take place at the same scale.

At $p_T \sim 1$~GeV/$c$ no perturbative scale stands far above the
hadronic ones. Collinear factorization drops corrections suppressed by
powers of the hadronic scale over the hard scale. Among them are the
ratios $m_h^2/p_T^2$ of the squared hadron mass to the squared transverse momentum, which are close to one for kaons
and baryons at these momenta. No controlled split into a computable
hard part and universal parton densities and fragmentation functions
exists there~\citep{collins2011foundations}.
Collins showed that factorization of diffraction, proved for deep-inelastic scattering, fails for hadron-hadron collisions~\citep{collins1998diffractive,collins1999erratum}.  The exchange mechanism considered in this review is a soft color singlet. Its excited systems are hadronic states, so every entry in the event record is a hadron.
The soft Pomeron is the Regge trajectory with the vacuum's quantum numbers.
It gives the soft total cross sections their slow rise~\citep{donnachie1992totalxsec}, and it dominates the fragmentation region. It is, however, not computed from first principles.

Analyticity, unitarity, and isospin are properties of the $S$-matrix and hold at every scale, soft included. Whether these principles, with Regge exchange and resonance duality, suffice to describe the soft spectra is the hypothesis this review sets out. The companion paper tests it against the NA49 data. Analyticity and unitarity constrain every amplitude through the finite-energy sum rules (Section~\ref{sec:duality}). Isospin relates charge states exactly, and Fischer and co-workers used it to separate target and projectile components in the NA49 data~\citep{chvala2004isospin,varga2003baryontransfer}. Regge exchange organizes the energy dependence~\citep{regge1959complex}. Its triple-Regge form organizes the fragmentation region (Section~\ref{sec:regge}). 

Resonance duality motivates a description of the excited systems in hadronic degrees of freedom. The resonance towers average to the Regge exchange that produces them, with the qualifications stated in Section~\ref{sec:duality}. These tools are statements about hadrons, the particles that the detectors record. The companion paper's model is built from them. Its fourteen fitted parameters are the quantities these statements leave unconstrained.

\section{Regge exchange and triple-Regge diffraction}\label{sec:regge}

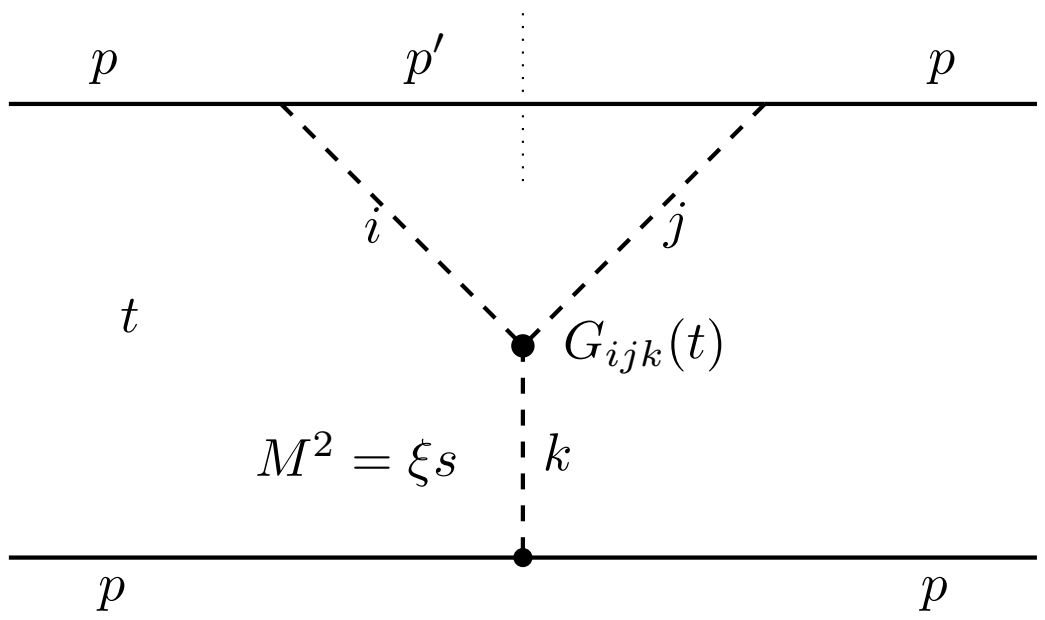
\begin{figure*}[htbp]
  \centering
  \begin{tikzpicture}[scale=1.0,>=latex]
    \draw[thick] (-3.4,2) -- (-1.6,2) node[pos=0.35,above] {$p$};
    \draw[thick] (-1.6,2) -- (1.6,2) node[pos=0.3,above] {$p'$};
    \draw[thick] (1.6,2) -- (3.4,2) node[pos=0.65,above] {$p$};
    \draw[thick,dashed] (-1.6,2) -- (0,0.4) node[midway,left] {$i$};
    \draw[thick,dashed] (1.6,2) -- (0,0.4) node[midway,right] {$j$};
    \filldraw (0,0.4) circle (2pt) node[right=4pt] {$G_{ijk}(t)$};
    \draw[thick,dashed] (0,0.4) -- (0,-1.0) node[midway,right] {$k$};
    \draw[thick] (-3.4,-1.0) -- (3.4,-1.0) node[pos=0.1,below] {$p$} node[pos=0.9,below] {$p$};
    \filldraw (0,-1.0) circle (1.6pt);
    \draw[thin,dotted] (0,2.6) -- (0,1.45);
    \node at (-2.6,0.6) {$t$};
    \node at (-1.1,-0.35) {$M^2 = \xi s$};
  \end{tikzpicture}
  \caption{Mueller's diagram~\citep{mueller1970} of a triple-Regge term in $p + p \to
  p' + X$, in the fragmentation region of the upper proton. The cross
  section is the discontinuity of the forward three-to-three amplitude
  $p + p + \bar p' \to p + p + \bar p'$ in the missing mass $M^2$ of
  the unobserved system $X$ (dotted cut), summed over $X$. The legs $i$ and $j$ carry the
  transfer $t$ between $p$ and $p'$. Their quantum numbers decide the
  leading line. With two Pomerons the leading proton stays a proton.
  The $k$ leg carries the sub-energy $\xi s$ of the excited system. A
  Pomeron there gives the triple-Pomeron term, and a Reggeon the
  $\mathbb{P}\mathbb{P}\mathbb{R}$ term.}
  \label{fig:triple_regge}
\end{figure*}

In the fragmentation region the triple-Regge expansion organizes the inclusive cross section. The expansion follows from the generalized optical theorem of \citet{mueller1970}. Figure~\ref{fig:triple_regge} shows the corresponding diagram. \citet{detar1971helicity} worked out the triple-Regge behavior near the boundary of phase space. \citet{goulianos1983diffractive} reviews its application to diffraction.
Let $\xi \equiv M^2/s$ be the fractional mass of the excited system. It equals $1 - x_F$ up to mass corrections. With $t$ the momentum transfer, the expansion at fixed $t$ reads
\begin{equation}\label{eq:triple_regge}
  \frac{d^2\sigma}{d\xi\,dt}
  = \sum_{i,j,k} G_{ijk}(t)\,
    \xi^{1 - \alpha_i(t) - \alpha_j(t)}\,
    (\xi s)^{\alpha_k(0) - 1} \,.
\end{equation}
The sum runs over the three exchanged trajectories. The $i$ and $j$ legs carry the momentum transfer $t$. The $k$ leg carries the excited system of sub-energy $\xi s$. The vertex function $G_{ijk}(t)$ describes the triple coupling.

The leading term is the triple-Pomeron term, $ijk = \mathbb{P}\mathbb{P}\mathbb{P}$.
\citet{donnachie1992totalxsec} give the Pomeron trajectory $\alpha_{\mathbb{P}}(t) \approx 1.0808 + 0.25\,t$. In the notation used here it reads $\alpha_{\mathbb{P}}(t) = 1 + \epsilon + \alpha'_{\mathbb{P}}\,t$, with $\epsilon = 0.0808$ and the slope $\alpha'_{\mathbb{P}} = 0.25$~GeV$^{-2}$.
Substituting the trajectory into Equation~\eqref{eq:triple_regge} gives
\begin{equation}\label{eq:ppp_t}
  \frac{d^2\sigma_{\mathbb{P}\mathbb{P}\mathbb{P}}}{d\xi\,dt}
  = G_{\mathbb{P}\mathbb{P}\mathbb{P}}(t)\, s^{\epsilon}\,
    \xi^{-1-\epsilon-2\alpha'_{\mathbb{P}} t}\,.
\end{equation}
In this form the exponent of $\xi$ depends on the transfer. Since $t < 0$, the factor $\xi^{-2\alpha'_{\mathbb{P}} t} = \exp[2\alpha'_{\mathbb{P}}\,t\,\ln(1/\xi)]$ steepens the $t$ distribution as $\xi$ falls. Take an exponential residue, $G_{\mathbb{P}\mathbb{P}\mathbb{P}}(t) \propto e^{b_0 t}$. The slope in $t$ is then $b_0 + 2\alpha'_{\mathbb{P}}\ln(1/\xi)$. This slope grows with the rapidity gap $\ln(1/\xi)$. Its growth is the shrinkage of the diffraction peak.

At $t = 0$ the mass law is a pure power. With $\xi = M^2/s$,
\begin{equation}\label{eq:ppp_law}
  \left.\frac{d^2\sigma}{dM^2\,dt}\right|_{t=0} \propto (M^2)^{-1-\epsilon}\,,
  \qquad \epsilon = \alpha_{\mathbb{P}}(0) - 1\,.
\end{equation}
Integrated over $t$, the law keeps a logarithmic factor. Take the exponential residue and integrate over $t$ up to $t \approx 0$. The integral is then $d\sigma/d\xi \propto \xi^{-1-\epsilon}/[b_0 + 2\alpha'_{\mathbb{P}}\ln(1/\xi)]$. At small $\xi$ it is flatter than Equation~\eqref{eq:ppp_law}. The pure power of Equation~\eqref{eq:ppp_law} is thus the zero-transfer form of the triple-Pomeron term. The integration over $t$ changes the mass dependence. A model that integrates over $t$ therefore needs its transfer law as well as its mass law.

The companion paper's model, HEART, takes the transfer dependence $\xi^{-1-\epsilon-2\alpha'_{\mathbb{P}} t}$ of Equation~\eqref{eq:ppp_t}. It uses both Donnachie--Landshoff values, $\epsilon = 0.0808$ and $\alpha'_{\mathbb{P}} = 0.25$~GeV$^{-2}$. In place of the residue it takes a vertex law $\exp(Bt)$. The masses of the two outgoing systems set the slope $B$. At each mass the vertex law is divided by its integral over the reachable transfer window. This per-mass normalization is a phenomenological prescription of the model. It replaces a mass-independent triple-Regge residue. Without the shrinkage, the model's mass law integrated over $t$ is the power of Equation~\eqref{eq:ppp_law}. With the shrinkage, that power is multiplied by the ratio of the transfer integrals with and without the shrinkage. At $t = 0$ the shrinkage factor equals one, so the joint density there carries only the power and the per-mass normalization. The shrinkage acts for $|t| \le 1$~GeV$^2$, the range in which the linear trajectory is fitted. Beyond that range the exponent is held at its value at $t = -1$~GeV$^2$. The protons measured at $x_F = 0.95$ set the slope of that tail of the transfer distribution.

The model holds $\epsilon$ at the Donnachie--Landshoff value, so that one soft-Pomeron intercept serves the whole model. Fitted free under the full law, $\epsilon$ is $0.027 \pm 0.031$. That value lies 1.8 standard deviations below 0.0808~\citep{heartpaper2026}.

The NA49 proton table ends at $x_F = 0.95$, that is, at $\xi \approx 0.05$. Leading protons above $x_F = 0.95$, resolved in $p_T$ in narrow $x_F$ bins, would measure the transfer slope against $\xi$ directly. Under the shrinkage of Equation~\eqref{eq:ppp_t} that slope grows by $2\alpha'_{\mathbb{P}}\ln(1/\xi)$. Between $\xi = 0.05$ and 0.005 the growth is about 1~GeV$^{-2}$. The same bins carry the low-mass end of the mass spectrum. There a $\mathbb{P}\mathbb{P}\mathbb{R}$ component would rise above the power law.

The triple-Regge form describes masses large against the hadronic scale and small against the collision energy, $1~\mathrm{GeV}^2 \ll M^2 \ll s$. At $\sqrt{s} = 17.3$~GeV this leaves masses from a few GeV$^2$ up to the coherence edge $M^2/s \approx 0.1$, about 30~GeV$^2$. Smaller masses lie in the resonance region of the excited proton. Taking $\xi$ toward zero at fixed $s$ moves the system into that region, where the triple-Regge form loses its validity.

Single-diffractive systems in HEART take their masses from the transfer-integrated law. The law applies inside the coherence region $M^2/s < 0.1$, from the $N\pi$ threshold up. Its continuation into the resonance region is a model prescription. In the sense of semilocal duality (Section~\ref{sec:duality}), the power law represents that region only on average.

The measured proton-dissociation cross section fixes the single-diffractive fraction of the events. Goulianos and Montanha list the values of Schamberger and collaborators at energies bracketing the SPS, namely 4.87~mb at $\sqrt{s} = 16.2$~GeV, 4.96~mb at 17.6~GeV, and $4.94 \pm 0.08$~mb at 19.1~GeV~\citep{goulianos1999factorization,schamberger1975}.
HEART takes $\sigma_{SD}(M^2/s < 0.05) = 4.9$~mb from these values.
Its systematic variants move this value by $\pm 0.8$~mb. That band covers a normalization uncertainty of about 10\% and the tension between the Fermilab and ISR measurements. It is wider than the uncertainty quoted for Schamberger and collaborators.

Subleading terms enter in two places, on the $k$ leg and on the $i$ and $j$ legs, with different quantum numbers.
In the $\mathbb{P}\mathbb{P}\mathbb{R}$ term the $k$ leg is an ordinary Reggeon. Its intercept is $\alpha_{\mathbb{R}}(0) = 0.5475$ in the Donnachie--Landshoff fit~\citep{donnachie1992totalxsec}. The $i$ and $j$ legs are Pomerons, so the leading proton stays a proton. The Reggeon describes the falling part of the Pomeron--proton cross section at the sub-energy $\xi s$. The $\mathbb{P}\mathbb{P}\mathbb{R}$ term is therefore a second component of the same single diffraction.
At equal $t$ its ratio to the triple-Pomeron term is $(G_{\mathbb{P}\mathbb{P}\mathbb{R}}/G_{\mathbb{P}\mathbb{P}\mathbb{P}})\,(\xi s)^{\alpha_{\mathbb{R}}(0)-\alpha_{\mathbb{P}}(0)}$. At fixed $\xi$ the ratio falls as $s^{-0.53}$. At fixed excited mass $M^2 = \xi s$ it does not depend on $s$. It grows toward low masses as $(M^2)^{-0.53}$. The intercept gap fixes this mass dependence. The ratio of the couplings fixes the normalization.

HEART's single-diffraction class has no $\mathbb{P}\mathbb{P}\mathbb{R}$ component. At the fitted parameters, a $\mathbb{P}\mathbb{P}\mathbb{R}$ term raises the $\chi^2$ of the six NA49 spectra, in a paired comparison on the same simulated events. The rise grows in proportion to its coupling. The spectra therefore take no additional low-mass single diffraction.

Non-vacuum exchanges on the $i$ and $j$ legs change the identity of the leading particle. There the isovector Reggeons $\rho$ and $a_2$ turn the leading proton into a neutron or a $\Delta$. They carry charge and isospin but no strangeness. The pion trajectory gives the one-pion exchange of the next paragraph. Strangeness exchange needs the $K^*$ and $K_2^*$ trajectories, which give a leading hyperon.

Reggeized one-pion exchange turns the leading proton into a neutron. Appleby and collaborators give its flux as a fixed pion term, in which the exchanged pion lies on a linear trajectory that vanishes at the pion pole~\citep{appleby2016pomeron},
\begin{equation}\label{eq:pionflux}
  \begin{split}
  \frac{d^2\sigma}{d\xi\,dt} &\propto
  \xi^{1-2\alpha_\pi(t)}\,\frac{-t}{(t-m_\pi^2)^2}\,S^2(\xi,q)\,,\\
  \alpha_\pi(t) &= 0.93\,(t-m_\pi^2)\,.
  \end{split}
\end{equation}
Here $\xi \approx 1 - x_F$ of the leading neutron. The factor $-t/(t-m_\pi^2)^2$ is the pion pole. It carries the two vertex couplings. The survival factor $S^2(\xi, q)$ accounts for the screening of the exchange by elastic scattering, with $q$ the transverse momentum of the leading neutron.
In the absorption model the Born amplitude is multiplied by the elastic $S$-matrix~\citep{sopkovich1962,gottfried1964absorption}.
Averaged over $t$, the survival factor falls with the leading neutron's momentum fraction $x_L = 1 - \xi$, from about one half at $x_L = 0.9$ to one third at $x_L = 0.6$. The companion paper multiplies the flux by the squared proton form factor and by the $\pi p$ total cross section at the sub-energy $\xi s$~\citep{donnachie1992totalxsec}.

In HEART the isovector exchange class is the $\rho$ and $a_2$ exchange at the leading lines. Its fitted fraction is zero, on the boundary of its range~\citep{heartpaper2026}. The one-pion-exchange class of HEART uses the flux of Equation~\eqref{eq:pionflux}.
The isovector Reggeon exchange and the pion exchange have different couplings. Their intercept gap alone therefore does not fix their relative size.

The quasi-exclusive end of the fragmentation region is $\xi \to 0$. There the intact proton has $x_F \to 1$. The excited mass then lies in the resonance region.

At fixed $\xi$ and over finite energy windows, the fragmentation region shows the approximate energy independence of limiting fragmentation (Section~\ref{sec:lf}). The corrections to it fall as $s^{-\Delta}$. Each exponent $\Delta$ is the gap between a subleading intercept and the Pomeron's in Equation~\eqref{eq:triple_regge}.
Pomeron dominance makes the cross section vary slowly with energy, as $s^{\epsilon}$ at fixed $\xi$ and $t$ (Equation~\eqref{eq:ppp_t}). Limiting behavior holds in that approximate sense. A density per inelastic event can cancel part of this rise against the rise of $\sigma_{\mathrm{inel}}$. That cancellation is a separate statement about the normalization.

At the other end, $\xi \to 1$, the excited mass approaches the full collision energy, $M^2 \to s$. The triple-Regge expansion no longer applies there, since the mass is no longer small against $\sqrt{s}$. Phase space and the residue cut off the growth of Equation~\eqref{eq:triple_regge}.

The perturbative descriptions of the Pomeron belong to the hard regime,
namely two-gluon exchange~\citep{low1975pomeron,nussinov1975puzzles}
and the summation of Balitsky, Fadin, Kuraev, and Lipatov
(BFKL)~\citep{kuraevlipatovfadin1977,balitskylipatov1978}.
They enter this review only as background. This review takes the soft
Pomeron of Equations~\eqref{eq:triple_regge}--\eqref{eq:ppp_law} from
the Donnachie--Landshoff fit. No first-principles calculation of it is
attempted here.

\section{Duality as the bridge}\label{sec:duality}

Resonance duality is the equivalence of two descriptions of one hadronic amplitude. One description is a sum over the $s$-channel resonances the reaction can form. The other is a sum over the $t$-channel Regge exchanges it can emit.
Each description continues into the other by analytic continuation.
The equivalence holds for averages of the amplitude over an energy window. As the energy rises, the resonances crowd together and overlap. Above some energy, no measurement resolves them one by one. The average that equals the Regge form is semilocal. It is taken over an energy window that is wide against the resonance spacing and narrow against the variation of the Regge form.

The finite-energy sum rules of \citet{dolenhornschmid1968} made the statement quantitative. In these sum rules, integrals over the resonance-dominated low-energy amplitude, determine the Regge parameters of the high-energy amplitude. In schematic moment form they read
\begin{equation}\label{eq:fesr}
  \int_0^{N} \nu^n\,\mathrm{Im}\,A(\nu,t)\,d\nu
  = \sum_i \beta_i(t)\,
    \frac{N^{\alpha_i(t)+n+1}}{\alpha_i(t)+n+1}\,.
\end{equation}
Here $\nu$ is the energy variable that changes sign under crossing of the $s$ and $u$ channels, $\nu = (s - u)/4m_N$ for pion-nucleon scattering. The upper limit $N$ lies at the end of the resonance region, above which the Regge form holds. The resonance-dominated left side fixes the Regge residues $\beta_i(t)$ and the trajectories $\alpha_i(t)$ on the right. The schematic form leaves out the specifics that a working sum rule fixes, namely the normalization of $A$, the moments $n$ allowed by its crossing symmetry, the threshold of the integral, and the Born or pole terms removed from it.

A Regge fit extrapolated down in energy reproduces the semilocal average of the resonance region. The two descriptions are therefore two forms of one amplitude. Adding them would count the same physics twice.

\citet{bloomgilman1970} found the same averaging rule in the lepton-proton system. There the deep-inelastic structure function follows the parton scaling curve once it is averaged over the resonance region. This parton-hadron duality shares the name because it shares the averaging rule. Resonance duality itself is stated entirely in hadronic degrees of freedom.

HEART assigns the two sides of the averaging rule to two stages of one event. The Regge side supplies the production kinematics, namely the masses, the momentum transfers, and the class fractions. The resonance side supplies the decay. No hadron is therefore generated twice.
The production masses and transfers follow the Regge law of Equation~\eqref{eq:ppp_t} with the model's vertex law (Section~\ref{sec:regge}). The decays follow the resonance tables of the Particle Data Group (PDG), in the 2024 edition, with the $N^*$ and $\Delta$ modes taken from the 2025 edition.

\citet{harari1968pomeranchuk} and \citet{freund1968fesr} split the amplitude into two components, each dual to one kind of exchange.
The imaginary part of a hadronic amplitude has two terms,
\begin{equation}\label{eq:twocomp}
  \mathrm{Im}\,A(s,t) = R(s,t) + P(s,t)\,.
\end{equation}
The resonance term $R$ is dual to ordinary $t$-channel Reggeon exchange.
The background term $P$ is dual to Pomeron exchange.

This two-component duality accounts for the pattern of the total cross sections. An exotic channel is one whose quantum numbers no $q\bar q$ meson or $qqq$ baryon can carry, such as $K^+p$ with baryon number one and strangeness $+1$. It forms no $s$-channel resonances. Duality therefore requires the ordinary Reggeon contributions to its imaginary part to cancel. Exchange degeneracy supplies the cancellation. It pairs the $\rho$ trajectory with the $a_2$ and the $\omega$ trajectory with the $f_2$, at common intercepts and residues. The near-constant $K^+p$ total cross section is the classic exotic case. Resonating channels such as $K^-p$ and $\bar p$+p fall with energy. Harari established this pattern across the measured pion-proton, kaon-proton, and proton-proton total cross sections.

Exchange degeneracy holds only approximately. The p+p channel is exotic too, yet the Donnachie--Landshoff fit keeps a Reggeon term in its total cross section~\citep{donnachie1992totalxsec}. That term is $56.08\,s^{-0.4525}$~mb, against $98.39\,s^{-0.4525}$~mb for $\bar p$+p. At $\sqrt{s} = 17.3$~GeV it is about 11\% of $\sigma_{\mathrm{tot}}$.

Duality motivates describing the excited systems in hadronic degrees of freedom. HEART builds that description at the event level, as a phenomenological model. Its decomposition into exchange classes, its continuum transitions, and its decay probabilities are additional model assumptions. The companion paper tests them against the NA49 data. Two-component duality constrains averages of amplitudes. It leaves these event-level choices open.

\citet{veneziano1968} supplied an explicit dual amplitude~\citep[see also][]{freund2007duality}. It uses the linear trajectories of Chew and Frautschi, $\alpha(s) = \alpha(0) + \alpha' s$~\citep{chewfrautschi1962},
\begin{equation}\label{eq:veneziano}
  A(s,t) = B(-\alpha(s),-\alpha(t))
  = \frac{\Gamma(-\alpha(s))\,\Gamma(-\alpha(t))}
         {\Gamma(-\alpha(s)-\alpha(t))}\,.
\end{equation}
Its poles at $\alpha(s) = 0, 1, 2, \ldots$ form the resonance tower. In the Regge limit of large $s$ at fixed $t$ the same amplitude behaves as Regge exchange. A tower of resonances and a Regge exchange are thus one object seen from the two channels.

HEART realizes the resonance side explicitly, as catalogued states plus a declared continuum density. It takes the Regge side as the production law. The string description of the same amplitude treats the scattering as that of one-dimensional objects, with the Pomeron as a closed string~\citep{freund2007duality}.

\citet{einhorngreenvirasoro1972} generalized the Harari--Freund conjecture from two-body amplitudes to inclusive reactions. In the inclusive form the secondary Reggeon terms fall with energy in the central plateau, where duality leaves the Pomeron alone. The fragmentation regions keep their subleading Reggeon terms.

Two terms in this review sound alike and are kept distinct. The \emph{two-component picture} of Fischer and co-workers is the experimental decomposition of soft spectra into a target component and a projectile component (Section~\ref{sec:lf}).
\emph{Two-component duality} is the Harari--Freund statement of this section.

The thesis of this review connects the two names as a hypothesis, namely that each experimental component is the decay of one duality object. Such an object is an excited system whose resonances, averaged over mass, reproduce the exchange that produced it. Duality motivates this hypothesis. The companion paper's fit to the NA49 spectra tests it.
Section~\ref{sec:precursors} traces the cascade description through its predecessors. Section~\ref{sec:heart} assembles the realization from the pieces of the earlier sections. Section~\ref{sec:measure} names the observables on which the two descriptions separate.

\section{Cascade predecessors}\label{sec:precursors}

Mueller's generalized optical theorem of 1970 is the formal backbone of every model in this section~\citep{mueller1970}. The ordinary optical theorem equates the total cross section to the imaginary part of the forward elastic amplitude. That imaginary part is a discontinuity, that is, the jump across the $s$-channel cut. Unitarity equates the discontinuity to the sum over all intermediate states.

Mueller generalized the construction to inclusive production. He crossed the detected hadron to the initial state. This turns $A + B \to h + X$ into the forward three-to-three amplitude $A + B + \bar h \to A + B + \bar h$. Forward means no momentum transfer between the initial and final triples. The discontinuity of this amplitude in the missing mass $M_X^2 = (p_A + p_B - p_h)^2$ sums over all unobserved $X$. That sum is the single-particle inclusive cross section. The inclusive spectrum is thus an elastic amplitude in disguise. Its Regge expansion is the triple-Regge form (Section~\ref{sec:regge}).

The models of this section differ in what they put inside the duality object. The candidates are a statistical fireball, a diffractively excited resonance cluster, and a cascade that conserves four-momentum and flavor at each vertex.

The nova model of \citet{jacobslansky1971,jacobslansky1972} is the closest named predecessor.
Novas are resonance-like objects produced diffractively with an energy-independent cross section. The cross section rises with the nova mass to a maximum below 2~GeV. Above that maximum, the cross section falls as Regge exchange and duality require. The novas decay mainly by cascade emission of pions. The model has no free parameters for the inclusive spectra. It still reproduces their shapes, the difference between meson and baryon distributions, the approach to ``scaling'', and the two-pion correlations observed at Argonne~\citep{jacobslanskywu1972,bergerohsmith1972}.
The 1972 papers also describe the proton, neutron, and $\Lambda$ spectra. They describe charge effects that survive at high energy as well.

HEART differs from the nova model in four places. It balances baryon number, charge, and strangeness at each vertex of the decay chain. It distinguishes single diffraction, double diffraction, and non-diffractive production. It fixes the normalization from measured cross sections. It writes an event record, so correlations beyond the single-particle spectra can be tested. The nova papers predict exclusive channels but generate no event records.

Hwa's 1971 diffractive model makes the cluster logic explicit~\citep{hwa1971diffractive}. In it, hadrons diffract into two clusters whose mean multiplicity is proportional to their mass,
\begin{equation}\label{eq:hwa_mult}
  \langle n \rangle \propto M\,.
\end{equation}
The single-particle spectrum then follows with no free parameter beyond the normalization. With that normalization, Hwa's model reproduces the measured inclusive spectrum. It specifies no production mechanism, because its clusters carry no exchange identity, no leading-particle structure, and no conservation chain.

HEART uses the proportionality as a test of its cascade. The companion paper compares the charged multiplicity of an excited system with its mass~\citep{heartpaper2026}. Below the open-charm threshold of 3.73~GeV, the cascade's count stands at 0.91 to 1.03 of the Mark~I multiplicity of $e^+e^-$ annihilation reactions~\citep{siegrist1982markI}.
Above that threshold, the count grows linearly with the mass. The measured multiplicity grows logarithmically.
Inside the measured range of Fermilab E-396, the charged multiplicity of the model's excited system lies 7\% to 17\% above $2(M_X - m_p)^{1/2}$~\citep{goulianos1982universality}.
The gap between the cascade's count and the measured laws thus grows inside the measured ranges.

Beyond those ranges the comparison is against the continued laws, namely the Mark~I logarithmic fit past 7.8~GeV and the E-396 square-root law past $M_X \approx 6$~GeV. At $M_X = 13$ to 17~GeV the cascade's count is 1.74 times the continued square-root law. In single diffraction a forward proton tags the excited mass. Measuring that mass together with the charged multiplicity of the recoiling system would decide between the cascade's count and the continued law. At the SPS the coherence region $M^2/s < 0.1$ ends at $M_X \approx 5.5$~GeV, inside the measured E-396 range. Diffractive masses of 13 to 17~GeV need $\sqrt{s}$ of about 55~GeV or more. They require a forward-proton tag at a collider.

\citet{goodwalker1960} supply the production side. In their picture the projectile is a superposition of components that the target absorbs differently, the diffractive eigenstates. Because the components are absorbed differently, the outgoing state differs from the incoming proton. That difference is the diffractive excitation. At each impact parameter the excitation cross section per unit transverse area is the variance of the scattering amplitude over the eigenstates. The cross section is its integral over the transverse plane,
\begin{equation}\label{eq:goodwalker}
  \sigma_{\mathrm{diff.\,exc.}} = \int d^2b\,
  \left[\langle |T(b)|^2 \rangle - |\langle T(b) \rangle|^2\right].
\end{equation}
Here $T(b)$ is the scattering amplitude of an eigenstate at impact parameter $b$. It is real for a purely absorptive amplitude. The brackets $\langle\cdot\rangle$ denote the average over the eigenstates in the incoming proton. Given Koba--Nielsen--Olesen scaling of the fluctuations, this picture is equivalent to triple-Regge diffraction~\citep{koba1972scaling,gustafson2013gw}. These are two  expressions for one excitation spectrum.

Neither formulation supplies the decay. The Good--Walker picture ends where the excited system is produced. The cascade models of this review begin there.
HEART's exchange layer anchors the single-diffraction fraction to the measured cross section, through the transfer-integrated mass law (Section~\ref{sec:regge}). It anchors the double-diffraction fraction to an estimated cross section. Both anchors take the place of a variance computed from a fluctuation model.

Hama's 1972 two-stage dissociation is the closest architectural ancestor~\citep{hama1972dissociation}.
In it the incident particle is first excited coherently as a whole. The excited state then decays independently of its production. This independence explains the observation that the mean multiplicity of the dissociated system does not change with the momentum transfer.
The structure is exchange followed by decay, as in HEART. Hama's decay is statistical, with no resonance table, no flavor tracking, and no conservation chain.

The two-fireball picture starts with Niu's two-center model of 1958 for cosmic-ray jets~\citep{niu1958fireball}. Each collision makes a forward and a backward fireball. The produced particles are emitted in groups from the two fireballs. \citet{fuchiniu1975} applied the model to multiple production at accelerator energies. \citet{detarsnider1970} showed that the phenomena behind the model, double diffraction among them, follow from multiperipheral mechanisms without fireballs.

The statistical fireball is the background that all cascade models argue against~\citep{hagedorn1965thermo,hagedorn1968ranft,frautschi1971bootstrap}.
A fireball assigns the excited system a ``temperature'' and lets it decay by available phase space. It describes the gross features of the spectra, such as the near-exponential fall of the transverse-mass distributions. Its basic form carries no leading-baryon quantum numbers and no resonance mass tails. Fireball implementations with exact conservation laws correlate flavors through those constraints, over the whole system. The cascade correlates them locally, at the hadronic vertex that creates them.

HEART keeps one element of the fireball picture, the Hagedorn growth of the density of states. Its continuum states above 2~GeV have the density
\begin{equation}\label{eq:hagedorn}
  \rho(M) \propto \exp(M/T_H)\,, \qquad T_H = 165~\mathrm{MeV}\,,
\end{equation}
with Hagedorn's limiting temperature $T_H$~\citep{hagedorn1965thermo,frautschi1971bootstrap}. These states decay by the two-body cascade described in the next two paragraphs, one hadron at a time.

The cascade alternative keeps the duality object hadronic at every step, namely resonances where they are resolved, a continuum above them, two-body transitions with measured branching fractions, and exact conservation at every vertex. The analyses of Fischer and co-workers already used this baseline. Fischer et~al.\ keep two-body decays to final hadrons in their direct counting, with three-body decays retained for the $\eta$ and $\omega$. They find about three quarters of inclusive $\pi^-$ from those decays, as a lower limit. They expect cascading to fill the remaining low-$p_T$, low-$x_F$ region~\citep{fischer2022piminus}. Varga's Quark Matter 2006 simulation keeps only two-body decays at measured branching fractions. It assigns its low-$p_T$ low-$x_F$ shortfall to neglected many-body decays and cascading~\citep{varga2006pppa}. Rybicki's thesis simulates the hadronic two-body decay of the parent resonance, at unit branching fraction in the studied channels~\citep{rybicki2003charged}.

HEART continues this baseline into the continuum. Catalogued resonances decay with their PDG branching fractions. A continuum state draws each step from a pool of two-body channels with the weight
\begin{equation}\label{eq:weights}
  w_c \propto C_{\rm iso}\,C_{\rm spin}\,q^*_c\;\gamma_s^{\,n_{s\bar s}}\;r_V^{\,n_V}\;r_{B\bar B}^{\,n_{B\bar B}}\;a_{YK}^{\,n_{YK}}\,.
\end{equation}
Here $C_{\rm iso}$ is the isospin Clebsch--Gordan factor, and $C_{\rm spin}$ counts the spin states of the established daughters. The factor $q^*_c$ is the two-body decay momentum of the channel. The four flavor factors weight a created strange pair ($\gamma_s$), a vector meson ($r_V$), a baryon pair ($r_{B\bar B}$), and an associated hyperon--kaon pair ($a_{YK}$). Each exponent counts how often its feature occurs in the channel. A channel either emits one established hadron and leaves a lighter continuum state, or it ends the chain in a pair of established hadrons~\citep{heartpaper2026}. Figure~\ref{fig:timeline} places these models in time.

\section{The event-level realization: HEART}\label{sec:heart}

\begin{figure*}[htbp]
  \centering
  \begin{tikzpicture}[>=latex]
    \tikzset{layer/.style={draw, thick, rounded corners,
      text width=2.6cm, minimum height=2.3cm, align=center,
      font=\small}}
    \node[layer] (ex) at (-4.8,0) {\textbf{exchange} \\
      \footnotesize p+p $\to A^\ast B^\ast$ \\
      \footnotesize $\mathbb{P}$: SD, DD, ND};
    \node[layer] (sp) at (-1.6,0) {\textbf{spectrum} \\
      \footnotesize $M,B,Q,S$ per system \\
      \footnotesize catalog / continuum};
    \node[draw, thick, rounded corners, minimum width=2.9cm,
      minimum height=2.3cm] (de) at (1.6,0) {};
    \node at (1.6,0.8) {\small \textbf{decay}};
    \filldraw (0.9,0.05) circle (1pt);
    \draw[thick] (0.9,0.05) -- (1.5,0.3) -- (2.0,0.42);
    \draw[thick] (1.5,0.3) -- (2.0,0.15);
    \node at (2.3,0.42) {\scriptsize $K^+$};
    \node at (2.3,0.15) {\scriptsize $\Lambda$};
    \draw[thick] (0.9,-0.55) -- (1.5,-0.55) -- (2.0,-0.3);
    \draw[thick] (1.5,-0.55) -- (2.0,-0.8);
    \node at (2.25,-0.3) {\scriptsize $p$};
    \node at (2.25,-0.8) {\scriptsize $\pi^0$};
    \node[layer] (ch) at (4.8,0) {\textbf{chemistry} \\
      \footnotesize continuum weights \\
      \footnotesize $\gamma_s, r_V, r_{B\bar B}, a_{YK}$};
    \draw[->,thick] (ex) -- (sp);
    \draw[->,thick] (sp) -- (de);
    \draw[->,thick] (de) -- (ch);
    \node[font=\footnotesize] at (0,-1.75)
      {Every vertex conserves four-momentum, baryon number, charge, and
      strangeness.};
  \end{tikzpicture}
  \caption{Schematic of one HEART event through the four layers. The
  exchange produces two excited systems, in single diffraction (SD),
  double diffraction (DD), or non-diffractive production (ND). The spectrum assigns each a
  mass and quantum numbers from the catalog or the continuum. The decay
  de-excites them by successive two-body transitions. The chemistry
  sets the four flavor weights of the continuum vertices. The
  catalogued states decay with their PDG branching fractions. Every
  vertex conserves four-momentum, baryon number, charge, and
  strangeness. The two branches are illustrative channels of two
  charge-one roots, ending as $K^+\Lambda$ and as $p\pi^0$. The figure
  shows the topology of an event and is not a simulated one.}
  \label{fig:heart_event}
\end{figure*}
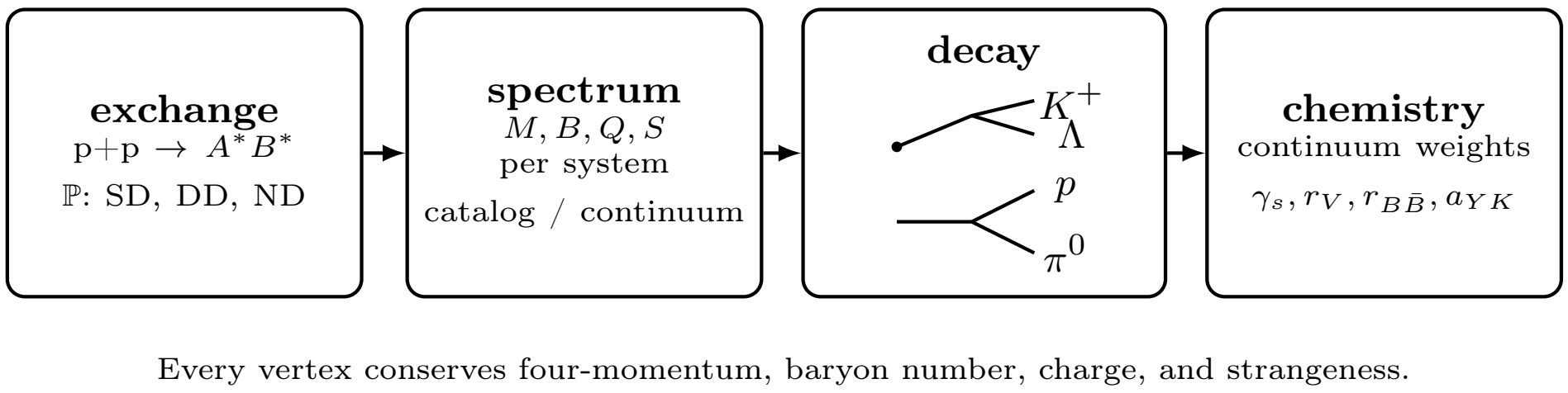

HEART is an event-level realization of the two-component picture. The companion paper defines the model and reports its fit~\citep{heartpaper2026}. This section shows how its four layers rest on the developments of the preceding sections. Figure~\ref{fig:heart_event} draws one schematic event through the layers.

The exchange layer assigns each inelastic event to one of four classes, namely single diffraction on either side, double diffraction, and non-diffractive production. Each class is a Regge exchange (Section~\ref{sec:regge}). Diffraction corresponds to the uncut Pomeron, an exchange that leaves a rapidity gap between the two systems. Its masses and transfers follow the triple-Pomeron law with the shrinkage of Equation~\eqref{eq:ppp_t}. The non-diffractive class is the cut Pomeron of Abramovsky, Gribov, and Kancheli, which excites both systems~\citep{abramovsky1973cuts}. The leading neutrons come from the reggeized pion exchange of Equation~\eqref{eq:pionflux}. The measured single-diffraction cross section and an estimated double-diffraction cross section~\citep{goulianos2004wherestand} fix the  fractions of single and double-diffractive event classes. The non-diffractive class takes the rest of the inelastic cross section.

The spectrum layer gives each excited system definite baryon number, charge, and strangeness. Below 2~GeV the system is a catalogued resonance, the resolved side of resonance duality (Section~\ref{sec:duality}). Above 2~GeV it is a member of a resonant continuum with the Hagedorn density of Equation~\eqref{eq:hagedorn}. The continuum stands for the unresolved resonances on average, in the semilocal sense.

The decay layer de-excites every system by successive two-body transitions through established hadrons. It continues the cascade lineage traced in Section~\ref{sec:precursors}. Every vertex conserves four-momentum, baryon number, charge, and strangeness. Catalogued states decay with their PDG masses, widths, and branching fractions. Continuum states draw their channels with the weight of Equation~\eqref{eq:weights}.

The chemistry layer is the four flavor factors of that weight, namely $\gamma_s$, $r_V$, $r_{B\bar B}$, and $a_{YK}$. They set the strangeness, vector, baryon-pair, and associated-production content of the continuum vertices.

The companion paper fits fourteen parameters to 1,254 points of the NA49 identified-hadron spectra and to the neutron $x_F$ distribution~\citep{na49pppi2006,na49ppk2010,na49ppproton2010}. The four flavor factors are among those parameters. The model has no separate central source. Its isovector exchange fraction fits to zero. The fit's $\chi^2$ is four to fourteen times lower than those of EPOS-LHC~\citep{eposlhc2015}, UrQMD~\citep{urqmd4manual2025}, and PYTHIA~8~\citep{pythia83guide} at their published tunes on the same points. The companion paper reports the fitted values and names the residual structure~\citep{heartpaper2026}.

A schematic event shows the mechanism in Figure~\ref{fig:heart_event}. A non-diffractive exchange produces a projectile-side system against a lower-mass target-side system. The projectile-side system carries the valence content $uud$ and a created $s\bar s$ pair. Its decay in Figure~\ref{fig:heart_event} has two branches. One branch is the $(Y, K)$ vertex of the leading baryonic line, which yields a $K^+$ and a $\Lambda$ by associated production. The companion paper names the momentum sharing at this vertex as the model's main limitation in the strange sector. The other branch ends as a proton and a neutral pion.

The $K^+$ and the $\Lambda$ are produced together at one vertex, so their strangeness balances locally by construction. A string description of the same final state creates the $s\bar s$ pair at a string breakup. Its fragmentation and remnant rules then set the pair's correlation. The fixed-target measurements proposed in Section~\ref{sec:measure} would show the strength and shape of that correlation in the pair's joint distribution in $x_F$ and $p_T$.

Table~\ref{tab:dictionary} translates between the Regge-era language
of this review and the generator language of the companion paper.

\begin{table*}[htbp]
  \centering
  \caption{Dictionary from Regge-era terms to event-generator terms.
  The cut-Pomeron row follows the cutting rules of Abramovsky, Gribov,
  and Kancheli~\citep{abramovsky1973cuts}. A cut exchange fills the
  rapidity gap and excites both systems. The gap is the empty interval
  in rapidity that a diffractive exchange leaves between the excited
  systems. An uncut exchange leaves the gap. It keeps one proton intact
  in single diffraction and excites both in double diffraction. The
  two Reggeon rows keep the separation made in Section~\ref{sec:regge}. The
  $\mathbb{P}\mathbb{P}\mathbb{R}$ term has a Reggeon on the $k$ leg
  and keeps the leading proton. An $f$ or $\omega$ exchange is the
  ordinary Reggeon of that term. Isovector Reggeons on the legs
  adjacent to the leading proton change its charge. Reggeized pion
  exchange is the one-pion-exchange class of the leading neutrons.
  Pionization is Mueller's name for central production that neither
  beam fragment carries~\citep{mueller1970}. Its row states how this review interprets it.}
  \label{tab:dictionary}
  \begin{tabular}{@{}lp{0.52\textwidth}@{}}
    \toprule
    Regge-era term & Generator term (HEART) \\
    \midrule
    Pomeron exchange & Color-singlet exchange class \\
    Cut Pomeron & Non-diffractive class (both systems excited) \\
    Diffractive system & Excited hadronic system of mass $M$ \\
    Triple-Pomeron coupling & $(M^2)^{-1-\epsilon-2\alpha' t}$ mass law, $\epsilon = 0.0808$, $\alpha' = 0.25$~GeV$^{-2}$ \\
    Pomeron--Pomeron--Reggeon & Low-mass single diffraction (no class) \\
    $\rho$, $a_2$ exchange at the leading line & Isovector exchange class \\
    Reggeized pion exchange & One-pion-exchange class, leading neutrons \\
    Duality object & Resonance or continuum member \\
    Fragmentation region & Decay products of one excited system \\
    Pionization, central plateau & Overlap of the two cascades \\
    \bottomrule
  \end{tabular}
\end{table*}

\section{Discriminating the cascade}\label{sec:measure}

The observables that separate a cascade from a string lie in the fragmentation region. Collider spectrometers with charged-hadron identification cover mid-rapidity. At $\sqrt{s} = 13$~TeV the projectile fragmentation region lies within about two units of $y_{\mathrm{beam}} = 9.5$. Only very-forward detectors measure there, namely the zero-degree calorimeters and the Roman-pot proton spectrometers. None of them identifies the produced charged hadrons.
At SPS energy the beam rapidity is $y_{\mathrm{beam}} = 2.9$, so the whole projectile fragmentation region lies in the forward hemisphere of one fixed-target spectrometer. With particle identification the NA49 p+p tables reach $x_F = 0.4$ for $K^+$ and $\bar p$, 0.5 for $K^-$, 0.85 for $\pi^-$, and 0.95 for protons. That acceptance is where the two descriptions can be distinguished. At $x_F \approx 0$ the two components overlap. There the two descriptions agree on the gross spectra. They disagree in the forward hemisphere, where the decay leaves its fingerprints. Three measurements would decide between them.

\begin{enumerate}
\item Identified forward baryons, with isospin and $p_T$ resolved,
test where the leading baryon's isospin is set. In the cascade the
forward neutrons come from the $N\pi$ decays of the excited proton
line and from the pion exchange, with no isospin parameter. A string
sets the forward $n/p$ ratio by the isospin of the diquark it picks
up. Neutron-tagged and proton-tagged samples at matched recoil mass
and transfer would separate the two.
\item The local association of $K^+$ and $\Lambda$, measured as a joint
$x_F$--$p_T$ distribution, tests the strength and kinematic structure
of associated strangeness. The HEART cascade (Section~\ref{sec:heart})
creates the pair at one hadronic vertex. A string produces compensating flavors next to each other where it breaks, by rules that differ between implementations. The test is therefore quantitative, a comparison of
predicted correlation functions.
\item Forward resonance signals, with their mass tails and the
alignment of the leading baryon's direction, test the two-body
transition structure directly against the fragmentation of a string.
The cascade assigns no spin or parity to the continuum, so hyperon
polarization lies outside its present predictions.
\end{enumerate}

Each observable lies at $x_F \gtrsim 0.2$, where the fragmentation structure is cleanest. The published identified-hadron tables become sparse there, most of all for the kaons. NA61/SHINE at the CERN SPS is the running instrument that can measure all three with particle identification. The case for precise soft-hadron measurements at this energy, given the geometry of collider detectors, is already stated in the 2022 global interpolation~\citep{fischer2022piminus}. These three measurements add the correlations that separate the two descriptions of the duality object.

\section{Conclusions}\label{sec:conclusions}

Fifty years after Benecke, Chou, Yang, and Yen, the fragmentation
regions still approach energy-independent limits. Fischer and co-workers interpret those limits in the NA49 spectra as the decay of
excited hadronic systems, a target component and a projectile
component. The tools this review applies to that picture predate QCD,
namely analyticity, unitarity, isospin, Regge exchange, and resonance
duality. Whether they suffice for the soft domain is the hypothesis the
review sets out.

Triple-Regge exchange organizes the fragmentation region. It supplies
the diffractive mass law, with the Pomeron's shrinkage in its transfer
dependence. Duality motivates describing the excited systems by towers of resonances whose average equals the exchange. The cascade description has predecessors from the nova model of Jacob and Slansky to Hama's
two-stage dissociation. HEART keeps their architecture of an exchange
followed by a decay.

The event-level realization, HEART, assembles these pieces in four
layers. Its exchange layer draws on triple-Regge diffraction, the cut
Pomeron, and reggeized pion exchange. The class fractions are anchored
to the measured single-diffraction cross section and an estimated
double-diffraction cross section. Its spectrum is catalogued resonances
below a duality boundary of 1.7 to 2.2~GeV and a Hagedorn continuum above it. Its decay is a cascade of
two-body transitions with exact conservation of four-momentum, baryon
number, charge, and strangeness. Four flavor factors weight the
continuum vertices.

Fourteen parameters fitted to 1,254 spectral points and the neutron
$x_F$ distribution describe the NA49 data. The model has no separate
central source. Its isovector exchange fraction fits to zero. The
fit's $\chi^2$ is four to fourteen times lower than those of three
generators at their published tunes on the same points. A companion
paper defines the model, reports the fit, and names the residual
structure.

The fit is consistent with the hypothesis up to the named residuals.
It does not separate the cascade from a string. The measurements that
would separate them lie in the fragmentation region, namely the isospin of forward baryons, the local balance of strangeness in associated production, and
forward resonance signals with their mass tails. Only a wide-acceptance
fixed-target spectrometer with particle identification measures there.

\subsection{Open questions}\label{sec:outlook}

Five questions remain open.
\begin{enumerate}
	\item The resonant continuum above 2~GeV is declared by the model. Its spectral density is fixed by construction, and its four flavor weights are fitted. A bootstrap would derive that density from the resonance spectrum instead. In Frautschi's sense the density is then fixed self-consistently by the spectrum it describes~\citep{frautschi1971bootstrap}.

	\item In the fitted model, associated strangeness, pair strangeness, and baryon--antibaryon production share one hadronic layer. That layer is HEART's chemistry layer (Section~\ref{sec:heart}). Pair-correlation measurements would test whether one layer can describe all three.

	\item The doubly strange baryons of the baryonic lines are open. The lines' $\Xi^-$ is estimated as the NA61/SHINE $\Xi^-$ yield less its $\bar\Xi^+$ yield, and the model's yield stands at 0.35 of that estimate. The pair-made $\bar\Xi^+$ stands at 1.46 of the NA61/SHINE measurement~\citep{na61xi2020}.

	\item The size of a charge-exchange Reggeon beside the pion flux is open. The neutron $x_F$ distribution tests the forward neutron-to-proton ratio. In the fit the neutrons follow the pion flux inside the calorimeter systematics. The isovector fraction fits to zero on its boundary~\citep{heartpaper2026}. A measurement that separates a charge-exchange Reggeon from the pion flux at fixed $x_F$ and $p_T$ would decide whether that ratio needs anything beyond the pion flux. The existing samples are the ISR neutron spectra at nonzero angles, resolved in $p_T$~\citep{engler1975neutron}, and the zero-degree ISR spectra at $p_T = 0$~\citep{flauger1976neutron}. Double-differential neutron cross sections exist in p+p at 12 and 24~GeV/$c$~\citep{blobel1978neutron}. The Fermilab neutrons from 400~GeV protons on hydrogen add a higher energy~\citep{whalley1979neutron}. The NA49 proton paper extracts $p_T$-integrated yields from the ISR and Fermilab sets. It concludes that those yields are not a basis for a quantitative comparison~\citep{na49ppproton2010}. The forward-neutron $dn/dx_F$ data of the Main Injector Particle Production experiment (MIPP) at Fermilab~\citep{mipp2011neutron} are integrated over $p_T$.

	\item The transfer to nuclei is open. It would repeat the same color-singlet exchange on each struck nucleon in p+C and p+Pb, with the nucleons sharing the light-cone energy. The p+p kernel would stay frozen throughout.
\end{enumerate}
Fixed-target measurements will decide which of these questions the data can answer.

\backmatter

\bmhead{Acknowledgments}
The NA49 Collaboration's published cross-section data and the global
interpolation of Fischer and collaborators are the data on which
this review stands. The author thanks both for making them public.

\section*{Statements and Declarations}
\bmhead{Funding}
No funding was received for conducting this study.

\bmhead{Competing interests}
The author has no relevant financial or non-financial interests to disclose.

\bmhead{Author contributions}
This review has a single author, who conceived the review, performed the literature search and the analysis, and wrote and revised the manuscript.

\bmhead{Data availability}
This manuscript has no associated data. [Author's comment: All data discussed are published in the cited papers.]

\bmhead{Code availability}
The HEART generator and its fit are kept at \url{https://github.com/ondrejch/SPS-HEART}.

\bmhead{Use of generative AI}
No generative AI tools were used to create or alter any images. Generative AI, primarily Anthropic Claude, was used for language editing, code, and analysis assistance during manuscript preparation. All technical content and figures were verified by the author.

\begin{appendices}
\section{Nomenclature and terminology}\label{sec:nomenclature}

\subsection{Kinematics and observables}

\paragraph{Invariant cross section.} $f = E\,d^3\sigma/dp^3$ is the
invariant cross section of a produced hadron species $h$. The NA49
tables give $f$. Divided by the inelastic cross section,
$f/\sigma_{\mathrm{inel}}$ is the invariant cross section per inelastic
event.

\paragraph{Feynman $x$.} $x_F \equiv 2p_L^*/\sqrt{s}$ is the
longitudinal scaling variable. Starred quantities are in the
center-of-mass frame. In the last bins at SPS energy the variable
$x_F' = p_L/p_L^{\max}$ replaces it.

\paragraph{Transverse momentum and energy.} $p_T$ is the transverse
momentum. $s$ is the squared center-of-mass energy.

\paragraph{Rapidity.} $y_h$ is the rapidity of a hadron of transverse
mass $m_T = \sqrt{m_h^2 + p_T^2}$. The beam rapidities are
$\pm y_{\mathrm{beam}}$, with $y_{\mathrm{beam}} \approx
\ln(\sqrt{s}/m_N)$. The beam-frame rapidity is $\eta^* \equiv y -
y_{\mathrm{beam}}$. The reduced rapidity is $y_{\mathrm{lab}} =
y_{\mathrm{beam}} - y = -\eta^*$. The projectile fragmentation region
is the interval of about two units below $\eta^* = 0$, that is, at
small $y_{\mathrm{lab}}$. At $\sqrt{s} = 17.3$~GeV the beam rapidity
is $y_{\mathrm{beam}} = 2.91$. Mid-rapidity then lies at $\eta^* =
-2.91$.

\paragraph{Masses.} $m_N$ is the beam nucleon mass, $m_\pi$ the pion
mass, and $M$ the mass of an excited hadronic system.

\subsection{Components and data treatment}

\paragraph{Two-component picture.} The decomposition of soft spectra
into a target component and a projectile component, each the decay of
one excited hadronic system, as Fischer and co-workers interpret the NA49 data. It is distinct from the two-component duality of Harari and
Freund.

\paragraph{Feed-over.} The target-projectile overlap. Each component
extends past mid-rapidity into a strip $\Delta\eta \sim 1$ wide.
Outside the strip each hemisphere holds one component. Its
fragmentation region lies within about two units of its beam rapidity.

\paragraph{Feed-down.} The subtraction of the products of weak decays,
namely the $V^0$ decays of $K^0_S$ and $\Lambda$, together with the $\Sigma$ and $\Xi$ decays that feed them. The published NA49 tables are feed-down
subtracted, since the $V^0$ yields were measured and their products
removed. Data without the subtraction are called feed-down included. HEART's
observables perform only the decays the tables keep
(Section~\ref{sec:lf}).

\paragraph{Central plateau, pionization.} The mid-rapidity region. In
this review it is the overlap of the two cascades.

\paragraph{Rapidity gap.} The empty interval in rapidity that a
diffractive exchange leaves between the excited systems.

\paragraph{The data.} The published NA49 cross-section data cover $0
\le x_F \le 0.95$ and $0.05 \le p_T \le 1.5$~GeV/$c$. The
non-perturbative sector is $p_T < 1.3$~GeV/$c$.

\paragraph{Multiplicity references.} Mark~I measured the mean charged
multiplicity of $e^+e^-$ annihilation from 2.6 to 7.8~GeV. It is the
reference for the continuum multiplicity. Fermilab E-396 measured the
multiplicity of diffractive dissociation, $\langle n \rangle \approx
2\,(M_X - m_p)^{1/2}$. It is the reference for the single-diffraction
systems.

\subsection{Hard scattering, for contrast}

\paragraph{Collinear factorization.} $Q$ is the hard momentum transfer
and $\Lambda_{\mathrm{QCD}}$ the QCD scale. $f$ denotes the parton
densities and $D$ the fragmentation functions. At $Q \sim 1$~GeV no
consistent split into a hard part and universal densities exists. The corrections suppressed by powers of $\Lambda_{\mathrm{QCD}}/Q$, which factorization drops, are then of order one.

\subsection{Regge exchange}

\paragraph{Trajectory.} $\alpha(t)$ is a Regge trajectory with
intercept $\alpha(0)$. The Pomeron carries the vacuum's quantum
numbers, with $\alpha_{\mathbb{P}}(t) \approx 1.0808 + 0.25\,t$.
Ordinary Reggeons have intercepts near $\alpha_{\mathbb{R}}(0) \approx
0.5$. Exchange degeneracy pairs the $\rho$ trajectory with the $a_2$
and the $\omega$ trajectory with the $f_2$.

\paragraph{Triple-Regge.} The organization of the fragmentation region
in $\xi \equiv M^2/s$ and $t$. The vertex function $G_{ijk}(t)$ couples
the three exchanged trajectories. The $i$ and $j$ legs carry the
transfer $t$ and decide the leading line. In the triple-Pomeron term
all three legs are Pomerons, $ijk = \mathbb{P}\mathbb{P}\mathbb{P}$.
In the $\mathbb{P}\mathbb{P}\mathbb{R}$ term the $k$ leg is a Reggeon,
and the leading proton stays intact.

\paragraph{Fractional mass $\xi$.} $\xi \equiv M^2/s$ is the fractional
mass of the excited system. It equals $1 - x_F$ up to mass
corrections. The product $\xi s$ is the sub-energy of that system. It
is the argument of the triple-Regge $k$ leg and of the $\pi p$ total
cross section in the pion flux. The limit $\xi \to 0$ is the
quasi-exclusive end, with $M^2 \ll s$ and an intact proton at $x_F \to
1$. The limit $\xi \to 1$ is $M^2 \to s$, where the triple-Regge form
no longer applies. In the one-pion-exchange flux $\xi \approx 1 - x_F$ for a leading neutron at $x_F$.

\paragraph{Diffractive mass law.} At $t = 0$ the law is
$d^2\sigma/dM^2\,dt \propto (M^2)^{-1-\epsilon}$, with $\epsilon =
\alpha_{\mathbb{P}}(0) - 1$. At $t < 0$ the triple-Pomeron exponent is
$-1-\epsilon-2\alpha'_{\mathbb{P}} t$. The extra term is the shrinkage
of the diffraction peak, with the slope $\alpha'_{\mathbb{P}} =
0.25$~GeV$^{-2}$. Corrections enter as $s^{-\Delta}$ at fixed $\xi$,
with $\Delta$ an intercept gap.

\paragraph{Exchange classes.} The four classes of inelastic events in
HEART, namely single diffraction (SD) on either side, double
diffraction (DD), and non-diffractive production (ND). The measured
$\sigma_{SD}(M^2/s < 0.05) = 4.9$~mb and an estimated $\sigma_{DD}$ fix
the class fractions. ND takes the rest of the inelastic cross section.

\paragraph{Leading baryons.} $x_L$ is the momentum fraction of the
leading baryon. The leading neutrons follow the reggeized
one-pion-exchange flux. Its survival factor $S^2(\xi, q)$ accounts for
the screening by elastic scattering, with $q$ the nucleon's transverse
momentum. Earlier data show structure in forward baryon production.
At the ISR the forward $\Delta^{++}$ and $\Lambda$ yields in p+p differ
from the $\bar\Delta^{--}$ and $\bar\Lambda$ yields in $\bar p$+p, as
quark counting with single $q\bar q$ annihilation
expects~\citep{chauvat1983leading}. In p+Pb at 19~GeV/$c$ the leading
baryons change flavor very differently from the expectation of simple
diquark-quark fragmentation of the incident
proton~\citep{barish2003leading}.

\paragraph{Isovector exchange.} The exchange of $I = 1$ Reggeons, the
$\rho$ and $a_2$ trajectories, at the leg adjacent to the leading
proton. It carries charge and isospin but no strangeness.

\paragraph{Good--Walker.} The projectile is a superposition of
diffractive eigenstates. The excitation cross section is the integral
over impact parameter of the variance of the eigenstate amplitudes
$T(b)$. Under Koba--Nielsen--Olesen scaling of the fluctuations it is
equivalent to triple-Regge diffraction.

\subsection{Duality and cascades}

\paragraph{Resonance duality.} One amplitude written as an $s$-channel
resonance sum or as a $t$-channel Regge sum. The two are equal for
semilocal averages, taken over energy windows wide against the
resonance spacing and narrow against the Regge variation. The
finite-energy sum rules of Dolen, Horn, and Schmid make the statement
quantitative. In them $\nu$ is the energy variable that changes sign under $s$--$u$ crossing, $N$ the upper
limit of the integral, and $\beta_i(t)$ the Regge residues.

\paragraph{Duality object.} The excited hadronic system that both sums
of a dual amplitude describe.

\paragraph{Two-component duality.} The Harari--Freund split
$\mathrm{Im}\,A = R + P$. The resonance term $R$ is dual to the
ordinary Reggeons, and the background $P$ is dual to the Pomeron.
Parton-hadron duality, found by Bloom and Gilman, is the same averaging
rule in the lepton-proton system.

\paragraph{Veneziano amplitude.} The crossing-symmetric Beta-function
amplitude. Its poles are the resonance tower, and its Regge limit is
Regge exchange.

\paragraph{Cascade precursors.} A nova is a diffractively produced
resonance-like object below 2~GeV that decays by pion cascade. Hwa's
clusters are diffractive pairs with $\langle n \rangle \propto M$. A
fireball decays statistically at a ``temperature''. A bootstrap fixes
the spectral density self-consistently from its own resonances.

\paragraph{Hagedorn density.} The density of hadronic states,
$\rho(M) \propto \exp(M/T_H)$, with the limiting temperature $T_H =
165$~MeV. HEART's continuum above 2~GeV carries this density.

\paragraph{Associated strangeness.} An $s$ and an $\bar s$, in a
hyperon $Y$ and a kaon $K$, created at one vertex of the decay chain.

\paragraph{Cut Pomeron.} In the cutting rules of Abramovsky, Gribov,
and Kancheli, a cut exchange fills the rapidity gap and excites both
systems. Table~\ref{tab:dictionary} translates this and the other
Regge-era terms into the generator terms of the companion paper.

\end{appendices}

\bibliography{references}

\end{document}